\documentclass{PoS}

\usepackage{amsmath}
\usepackage{amsfonts}
\usepackage{amssymb}

\usepackage{cleveref}

\usepackage{xspace}
\usepackage{macros}

\title{First results from an extended freed-isobar analysis at COMPASS}

\ShortTitle{First results from an extended freed-isobar analysis at COMPASS}

\author{\speaker{Fabian Michael Krinner}%
         \thanks{For the COMPASS collaboration}\\
        Institute for Hadronic Structure and Fundamental Symmetries\\
        E-mail: \email{fabian-krinner@mytum.de}}

\abstract{One of the goals of the COMPASS experiment is the precision study of 
light-meson spectroscopy with data for various final states. With $46\times
10^6$ exclusive events, the process \process constitutes the flagship channel.

Based on this data set, the results of an extensive partial-wave analysis (PWA),
 using a total of 88 partial waves in the model, were published. Along with it, 
results of a first study of the so-called freed-isobar method were shown. Here, 
the fixed amplitudes for appearing $\Ppip\Ppim$ intermediate states used in the 
conventional analysis were replaced by sets of piecewise constant functions to 
extract the amplitudes of the $\Ppip\Ppim$ subsystems directly from the data. In
 this first study, this was done for three $\Ppi$ partial waves with $\JPC= 
0^{++}$ intermediate $\Ppip\Ppim$ states.

The promising results inspired further extension of this method, by also 
freeing intermediate $\Ppip\Ppim$ states with $\JPC = 1^{--}$ and $2^{++}$. 
With this extension of the set of freed waves, mathematical ambiguities in the 
model arise due to the much higher flexibility of the model. We will present 
first results of these extended studies on COMPASS data along with methods to 
resolve the arising ambiguities.}

\FullConference{XVII International Conference on Hadron Spectroscopy and Structure\\
                 25-29 September, 2017\\
                 University of Salamanca, Salamanca, Spain}
\begin{document}

\section{Freed-isobar partial wave analysis}
\label{sec::introduction}
Due to the large data set collected by COMPASS, more elaborate analysis methods 
become applicable. One of these methods is the so called freed-isobar PWA. This 
approach is an extension of the conventional three-body PWA, that relies on the 
isobar model. In this model, the process is modeled as a two step decay of an 
intermediary state $\PX$ into one of the final state \Ppim and another 
intermediary state \Pxi, the isobar. The isobar subsequently decays into \Ppip
\Ppim. Within this model the intensity distribution $\intens\lrps$ for the 
$3\Ppi$ final state is modeled as coherent superposition of waves with given 
angular momentum quantum numbers and known isobar resonances. These quantum 
numbers are the spin \Jj and eigenvalues under parity and generalized 
charge conjugation---\PP and \Cc---, the spin projection \Mm, and reflectivity 
\eps of \PX, as well as the analogous quantum numbers $\JPC_\Pxi$ of the isobar.
With these quantum numbers defining a partial wave, we can write the intensity 
distribution:
\begin{equation}
\intens\lrps=\intensAbs{\sum_{i\,\in\,\text{waves}}\prodAmp_i\left[\angular_i\lrps\shape_i\lr{\mTwoPi} + \boseText\right]}
,\end{equation}
as a function of the phase space variables \ps. The complex-valued production 
amplitudes $\prodAmp_i$ describe the intensities and relative phases of the 
partial waves, the angular amplitudes $\angular_i\lrps$ describe their angular 
distributions, and the dynamic amplitudes $\shape\lr{\mTwoPi}$ describe the 
dependence of the amplitudes of the $\Ppip\Ppim$ system on \mTwoPi. The 
production amplitudes are free parameters in a likelihood fit to the measured 
\ps distribution and the angular amplitudes are fully determined by the 
appearing angular-momentum quantum numbers. However, the dynamic amplitudes are 
not determined by first principle, but are a necessary model input; for the most
 simple case of isolated $\Ppip\Ppim$ resonances, Breit-Wigner amplitudes are 
commonly used. Due to computational reasons, the dynamic amplitudes cannot 
have any free fit parameters.

To avoid the necessity for this model input in the freed-isobar approach, 
the dynamic amplitudes are replaced by sets of step-like functions, that 
approximate the dynamic amplitude by a set of constants $\binVal^b$ within 
narrow bins $b$ of \mTwoPi \cite{first::fi}:
\begin{equation}
\shape_i\to\sum_{b\,\in\,\text{bins}}\binVal_i^b\step^b\lr{\mTwoPi},\quad\text{where}\quad\step^b\lr{\mTwoPi}=\begin{cases}1 & \text{if \mTwoPi in $b$},\\
    0 & \text{otherwise}.
   \end{cases}
\end{equation}
In this approach, the step-like functions appear like ``normal'' partial waves 
in the fit. This allows us to use the same likelihood method as in the 
conventional PWA and determine the dynamic isobar amplitudes encoded by the 
free parameters $\binVal^b$ directly from the data.

Such a set of steplike functions will be denoted will be denoted by 
$\LR{\Ppi\Ppi}_{\JPC_\Pxi}$, where $\JPC_\Pxi$ are the \JPC quantum numbers of 
the freed isobar. Partial waves with steplike dynamic isobar amplitudes will be 
called freed-isobar  waves, waves with fixed parameterizations of the dynamic 
isobar amplitudes will be called fixed-isobar waves. Within the freed-isobar 
approach, models with any combination of freed- and fixed-isobar waves are 
possible.

\section{Freed-isobar PWA by COMPASS}
\label{sec::oldStudy}
A first freed-isobar analysis based on an extensive PWA of a data set of 
$46\times10^6$ events for the process $\process$ collected by COMPASS in 2008 was
 published in ref.~\cite{big::compass} using a set of 88 waves. In this first 
analysis, seven fixed-isobar waves are replaced by three freed-isobar ones, due 
to matching angular momentum quantum numbers:
\begin{equation}\begin{aligned}
&0^{-+}0^+\PfZeroS\Ppi\Sw,\ 0^{-+}0^+\PfZeroN\Ppi\Sw,\ 0^{-+}0^+\PfZeroF\Ppi\Sw\ &\to&\ 0^{-+}0^+\LR{\Ppi\Ppi}_{0^{++}}\Ppi\Sw\\
&1^{++}0^+\PfZeroS\Ppi\Pw,\ 1^{++}0^+\PfZeroN\Ppi\Pw\ &\to&\ 1^{++}0^+\LR{\Ppi\Ppi}_{0^{++}}\Ppi\Pw\\
&2^{-+}0^+\PfZeroS\Ppi\Dw,\ 2^{-+}0^+\PfZeroN\Ppi\Dw\ &\to&\ 2^{-+}0^+\LR{\Ppi\Ppi}_{0^{++}}\Ppi\Dw
.\end{aligned}\end{equation}
This first study yielded very promising results, that helped understand the 
$\LR{\Ppi\Ppi}_{0^{++}}$ wave and to eliminate the possibility that the \PaOneF 
resonance-like signal, first observed by COMPASS in ref.~\cite{a1::1420}, is an 
artifact of the parameterization of the \PfZeroN dynamic isobar amplitude. 
Inspired by these results, we made attempts to extend the freed-isobar approach 
to a larger set of partial waves.

\section{Ambiguities in the freed-isobar approach}
\label{sec::freedIsobar}
With increasing number of freed-isobar waves, the flexibility of the analysis 
model drastically increases, since one complex-valued free parameter appears for 
every \mTwoPi bin of every freed-isobar wave. For some combination of freed 
waves, this flexibility allows to approximate continuous ambiguities in the 
analysis model, that have to be identified an resolved. Such ambiguities will be
 called zero modes in this article. A more detailed discussion of zero modes can
 be found in ref.~\cite{freed::isobar}.
\subsection{Appearance of zero modes}
Since in the freed-isobar approach, the dynamic isobar amplitudes of freed 
waves can take any shape---up to the finite \mTwoPi bin width---the zero modes 
correspond to specific shapes $\zero_i\lr{\mTwoPi}$ of these dynamic amplitudes, 
that cancel exactly to zero at evert point \ps in phase space:
\begin{equation}
\sum_{i\,\in\,\text{waves}}\angular_i\lrps\zero_i\lr{\mTwoPi}+\boseText = 0
,\end{equation}
where the sum runs over all waves that are affected by the particular zero mode.
 Since the cancellation is exact, a shift of the isobar dynamic amplitudes 
$\shape_i\lr{\mTwoPi}$ by these shapes $\zero_i\lr{\mTwoPi}$ does not change 
the overall amplitude and thus the intensity. The $\zero_i\lr{\mTwoPi}$ 
constitute a continuous ambiguity in a freed-isobar analysis:
\begin{equation*}
\intensAbs{\sum_{i\,\in\,\text{waves}}\angular_i\lrps\shape_i\lr{\mTwoPi}+\boseText}=\intensAbs{\sum_{i\,\in\,\text{waves}}\angular_i\lrps\LR{\shape_i\lr{\mTwoPi}+\zeroCoeff\zero_i\lr{\mTwoPi}}+\boseText}
.\end{equation*}
This ambiguity is represented by the arbitrary complex-valued coefficient
\zeroCoeff.
\subsection{Resolution of zero modes}
\label{sec::zeroModeResolution}
The presence of zero modes may affect the result of a freed-isobar PWA and has 
to be accounted for, since the extracted dynamic isobar amplitudes 
$\binVal^{\text{fit},b}$ can be shifted away from the physical ones 
$\binVal^{\text{phys},b}$:
\begin{equation}
\binVal^{\text{phys},b}_{i}=\binVal^{\text{fit},b}_i+\zeroCoeff\binVal^{0,b}_i
,\end{equation}
where $\binVal^{0,b}_i$ is the shape $\zero_i\lr{\mTwoPi}$ of the zero mode for 
wave $i$ evaluated at the \mTwoPi bin $b$.

To remove the effect of the zero mode, we have to determine \zeroCoeff such, 
that the resulting combination 
$\binVal^{\text{fit},b}_i+\zeroCoeff\binVal^{0,b}_i$ resembles the physical 
solution. This is done, by performing a second fit step, where the following 
$\chi^2$ function is minimized:
\begin{equation}\label{eqn::chi2}
\chi^2\lr{\zeroCoeff}=\sum_{i,b}\sum_{j,c}\delta_i^b\lr{\zeroCoeff}\lr{\mathbf{C}^{-1}}_{i,j}^{b,c}\delta_j^c\lr{\zeroCoeff}
,\end{equation}
where $\mathbf{C}_{i,j}^{b,c}$ is the covariance matrix of the 
$\binVal^{\text{fit},b}_i$ obtained by the minimizing algorithm and the 
differences $\delta_i^b\lr{\zeroCoeff}$ are given by:
\begin{equation}
\delta_i^b\lr{\zeroCoeff} = \binVal^{\text{fit},b}_i+\zeroCoeff\binVal_i^{0,b}-\binVal_i^{\text{model},b}
.\end{equation}
The values $\binVal_i^{\text{model},b}$ are the values of a model for the 
dynamic isobar amplitude of wave $i$ evaluated at the corresponding \mTwoPi bin.
Such a model can, for example, be a Breit-Wigner amplitude, if the 
corresponding wave is known to be dominated by a single resonance, as is true for 
the \PrhoM waves with $\LR{\Ppi\Ppi}_{1^{--}}$ isobar, or for the \PfTwoM isobar 
for waves with $\LR{\Ppi\Ppi}_{2^{++}}$ isobar.

\begin{figure}
  \includegraphics[width=.32\textwidth]{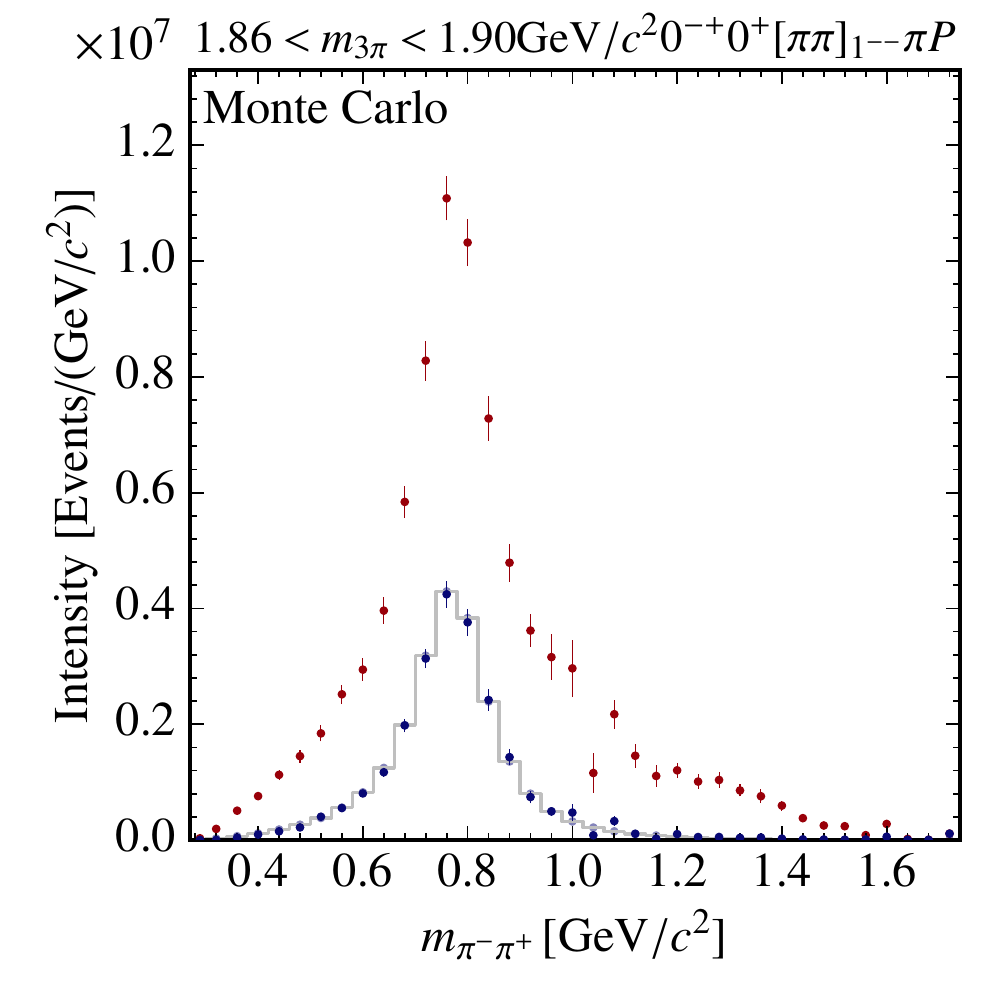}
  \includegraphics[width=.32\textwidth]{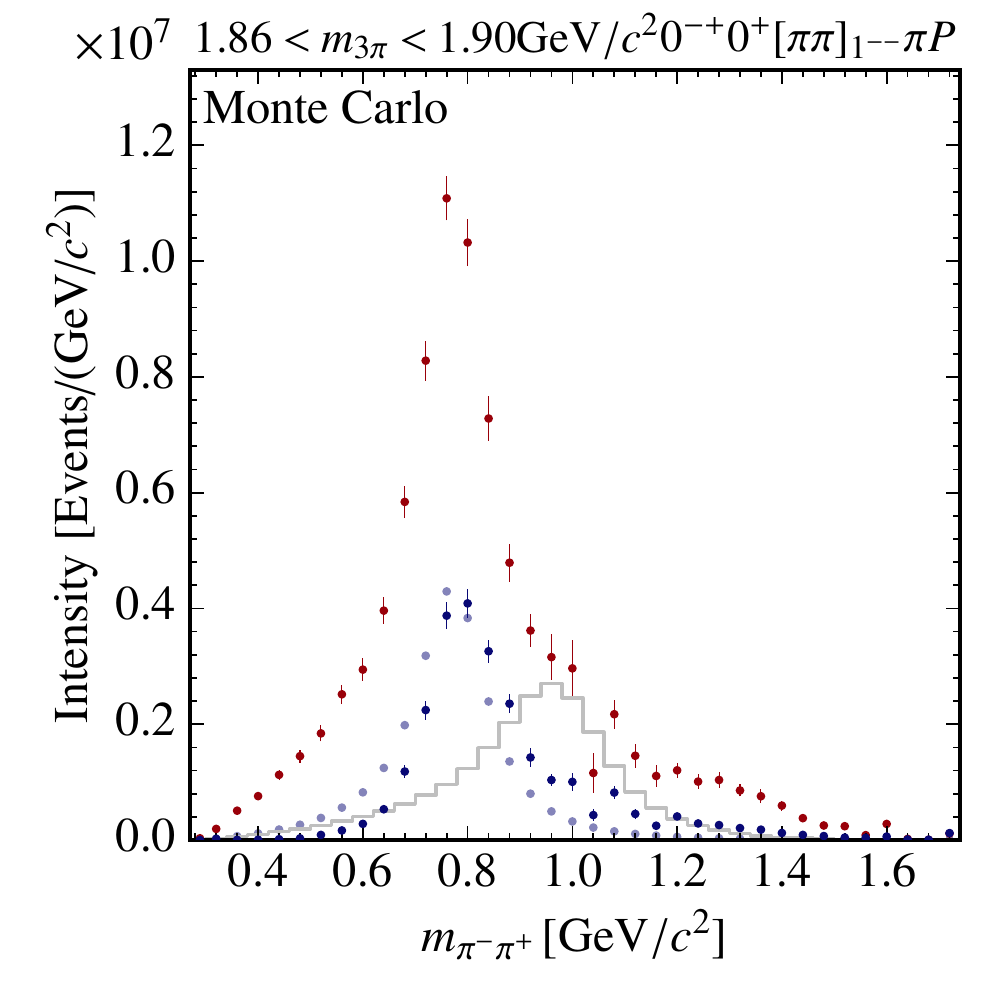}
  \includegraphics[width=.32\textwidth]{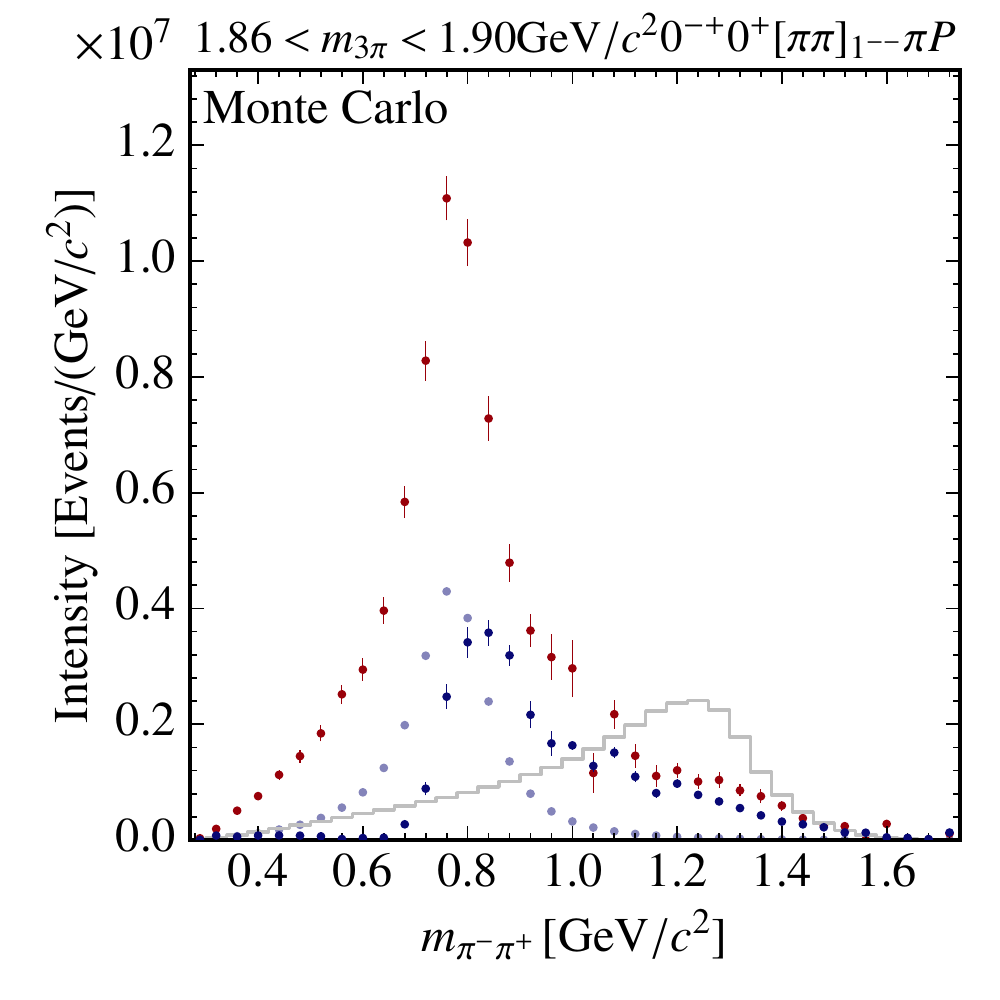}
\caption{Results of the freed-isobar fit before (red) and after resolving of the
         zero mode ambiguity (blue) with different constraints (gray line). The 
         shape used to generate the Monte Carlo data is indicated by light blue 
         points.\label{fig::wrongFixing}}
\end{figure}

This second fit adjusts only one complex-valued parameter, 
while the freed-isobar fit extracts a complex-valued parameter for every \mTwoPi
 bin. Therefore, most of the information obtained with the freed-isobar method 
is kept, even in this second fit step is performed. To demonstrate, that 
resolving zero-mode ambiguities cannot change the 
resulting dynamic isobar amplitudes arbitrarily, we generated a Monte Carlo data
 set, analyzed it with the freed-isobar method, and resolved the zero-mode 
ambiguities using a Breit-Wigner parameterization with three different values 
for mass and width as constraints. The results of this study are shown in 
\cref{fig::wrongFixing}. It can be seen, that the resulting position of the 
\PrhoM peak (blue points) is nearly independent of the chosen constraint (gray 
line) and that using the 	correct constraint yields the Monte Carlo input shape.
\subsection{Example: spin-exotic wave}
\label{sec::spinexotic}
One example for the appearance of a zero mode is the freed spin-exotic $1^{-+}
1^+\LR{\Ppi\Ppi}_{1^{--}}\Ppi\Pw$ wave. In this case, the zero mode is contained
 within this single wave. Therefore, the shape of the zero-mode has to fulfill:
\begin{equation}\label{eqn::exoticZeroCondition}
\angular_{1^{-+}}\lrps\zero_{1^{-+}}\lr{\mTwoPi}+\boseText=0
.\end{equation}
The angular amplitude for the spin-exotic wave is given by:
\begin{equation}
\angular_{1^{-+}}\lrps\propto\vec p_1\times\vec p_3
,\end{equation}
where $\vec p_{1,3}$ are the three-momenta of the two \Ppim in the final state 
in the rest frame of the three-pion system. Inserting this into 
\cref{eqn::exoticZeroCondition} and using the antisymmetry of the cross product,
 we obtain:
\begin{equation}\label{eqn::exoticZeroCondition2}
0=\zero_{1^{-+}}\lr{\mTwoPi}\vec p_1\times \vec p_3 + \zero_{1^{-+}}\lr{\mTwoPiBose}\vec p_3\times \vec p_1 = \vec p_1\times\vec p_3\LR{\zero_{1^{-+}}\lr{\mTwoPi}-\zero_{1^{-+}}\lr{\mTwoPiBose}}
,\end{equation}
where \mTwoPiBose is the Bose symmetrized invariant two-pion mass with 
final-state particles 1 and 3 interchanged. \Cref{eqn::exoticZeroCondition2} is 
fulfilled, if $\zero\lr{\mTwoPi}$ is a constant, which is the shape of the 
zero mode in this wave.

To resolve the ambiguity caused by this zero mode we use the method introduced 
in \cref{sec::zeroModeResolution}, using a Breit-Wigner parameterization with 
fixed mass and width for the \PrhoM resonance as constraint. To check 
consistency we try a second approach, where we leave the mass and width of this 
parameterization as additional free parameters in the minimization of 
\cref{eqn::chi2}. As final value for \zeroCoeff, we use the weighted average of 
both approaches. 

\section{Extended freed-isobar PWA}
\label{sec::newStudy}
Based on the first freed-isobar analysis published in ref.~\cite{big::compass}, 
we performed an extended freed-isobar analysis with twelve freed waves:
\begin{equation}\begin{aligned}
0^{-+}0^+\LR{\Ppi\Ppi}_{0^{++}}\Ppi\Sw\quad &\ 1^{++}0^+\LR{\Ppi\Ppi}_{1^{--}}\Ppi\Sw\ &\ 2^{-+}0^+\LR{\Ppi\Ppi}_{0^{++}}\Ppi\Dw\quad &\ 2^{-+}0^+\LR{\Ppi\Ppi}_{2^{++}}\Ppi\Sw\\
0^{-+}0^+\LR{\Ppi\Ppi}_{1^{--}}\Ppi\Pw\quad &\ 1^{++}1^+\LR{\Ppi\Ppi}_{1^{--}}\Ppi\Sw\ &\ 2^{-+}0^+\LR{\Ppi\Ppi}_{1^{--}}\Ppi\Pw\quad &\ 2^{-+}1^+\LR{\Ppi\Ppi}_{1^{--}}\Ppi\Pw\\
1^{++}0^+\LR{\Ppi\Ppi}_{0^{++}}\Ppi\Pw\quad &\ 1^{-+}1^+\LR{\Ppi\Ppi}_{1^{--}}\Ppi\Pw\ &\ 2^{-+}0^+\LR{\Ppi\Ppi}_{1^{--}}\Ppi\Fw\quad &\ 2^{++}1^+\LR{\Ppi\Ppi}_{1^{--}}\Ppi\Dw
.\end{aligned}\end{equation}
To minimize potential leakage effects 11 of the freed waves were chosen to be 
the eleven waves with the larges intensity in the conventional analysis in 
ref.~\cite{big::compass}. In addition, the spin-exotic $1^{-+}1^+\PrhoM\Ppi\Pw$ 
wave was freed, since it is a wave of major interest in the 
analysis. With this choice of freed waves, the analysis model has 12 fixed- and 
72 freed-isobar waves. Due to the high flexibility of this model, five zero modes 
appear: One in the $\JPC\Mrefl=0^{-+}$ sector, one in the $1^{++}0^+$ sector,
 two in the $2^{-+}0^+$ sector, and one on the spin-exotic wave, as discussed in
 \cref{sec::spinexotic}. In this article, we only show results for the 
spin-exotic wave, therefore only the zero mode in this wave has to be 
resolved, which we do as described in \cref{sec::spinexotic}. 

\begin{figure}
  \includegraphics[width=.5\textwidth]{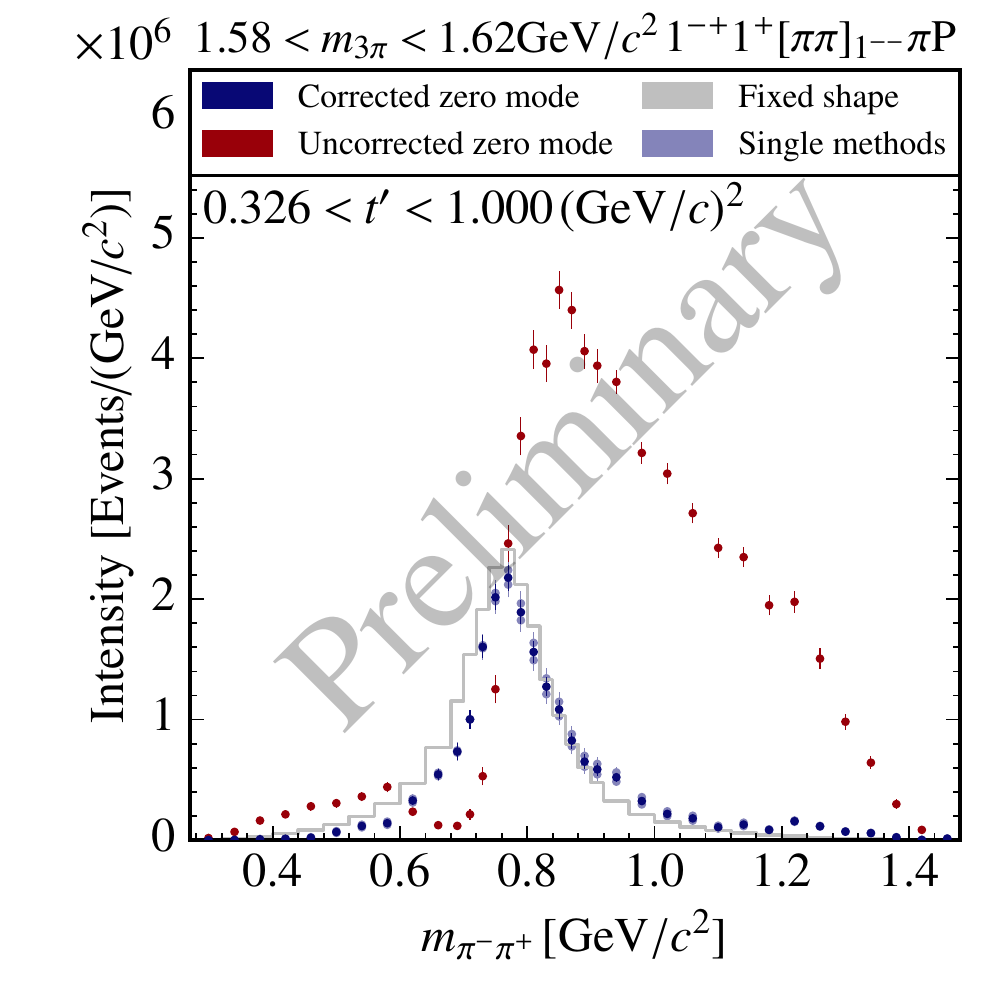}
  \includegraphics[width=.5\textwidth]{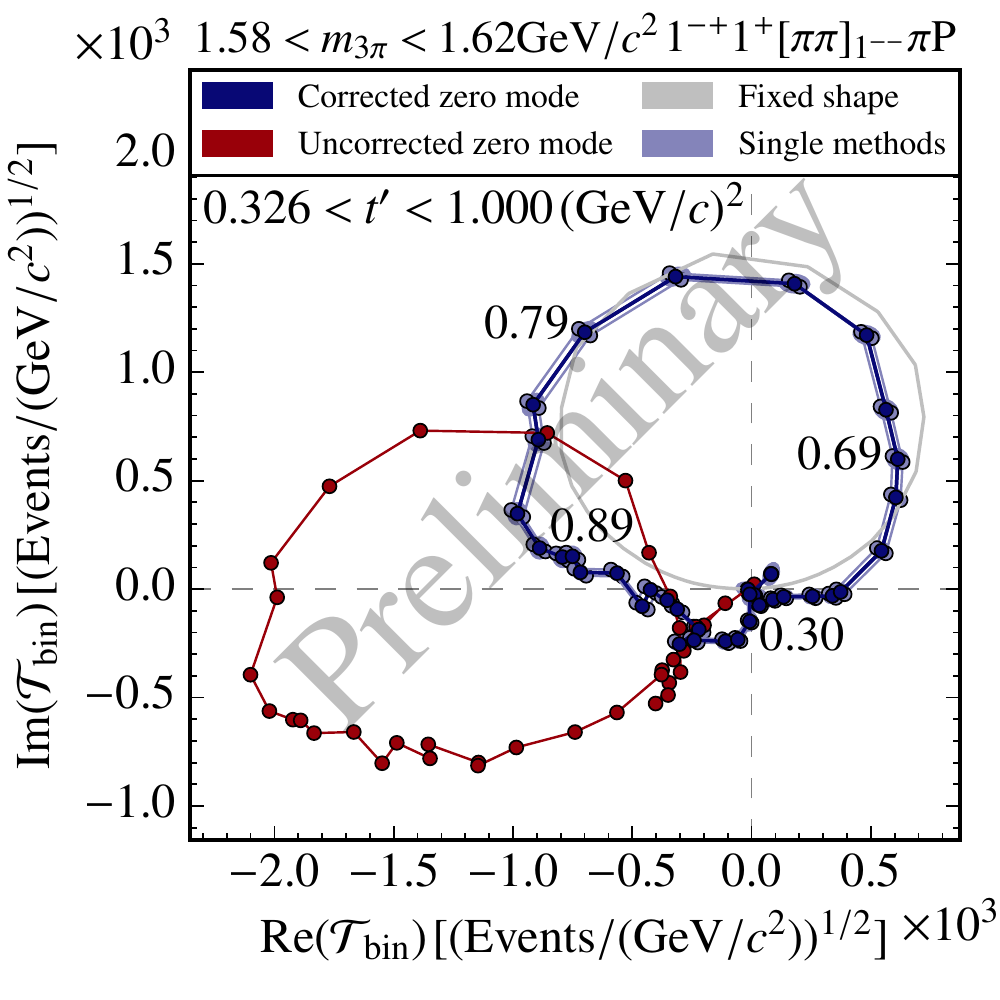}
\caption{Intensity distribution and Argand diagram for the dynamic-isobar 
         amplitude for the $1^{-+}1^+\LR{\Ppi\Ppi}_{1^{--}}\Ppi\Pw$ wave in a
         bin of the three-pion mass and the squared four momentum transfer 
         $t^\prime$.\label{fig::result}}
\end{figure}

The result for the dynamic isobar amplitude for the spin-exotic wave is shown 
in \cref{fig::result}. The resulting dynamic isobar amplitude is dominated by 
the \PrhoM resonance, which justifies the use of this resonance as constraint 
for resolving the zero-mode ambiguity. This is also supported by the fact, that 
resolving the ambiguity with floating \PrhoM mass and width yields compatible 
results. These are shown in light blue in \cref{fig::result}.

Even though the dynamic isobar amplitude is dominated by the \PrhoM 
resonance, we find, that the exact shape differs significantly from a pure 
Breit-Wigner shape, since the low mass tail rises slower and the high mass tail 
falls slower, than a pure Breit-Wigner amplitude. Reasons for this behavior can 
be rescattering effects with the third pion in the final state or non-resonant 
contributions to the process. 

\section{Conclusions and outlook}
\label{sec::conclusion}
We performed a freed-isobar analysis of COMPASS data for the \process channel, 
based in the analysis published in ref.~\cite{big::compass}, using an extended 
set of 12 freed waves. 
In this analysis model, we identified continuous ambiguities---zero modes---and 
showed a way to resolve them using additional constraints on the dynamic isobar 
amplitudes. 
With this method, we were able to extract the dynamic isobar amplitude for the 
spin-exotic $1^{-+}1^+\LR{\Ppi\Ppi}_{1^{--}}\Ppi\Pw$ wave from the data, shown 
in \cref{fig::result}. Even though this dynamic isobar amplitude is dominated by
 the \PrhoM resonance, it shows significant deviations from a simple 
Breit-Wigner shape.

As next steps, we will extend the analysis to all bins in the three-pion 
mass and the squared four-momentum transfer. We will also free the dynamic 
isobar amplitudes of other interesting waves, for example in the $\JPC=3^{++}$ 
or $4^{++}$ sectors.

\end{document}